\documentstyle[psfig]{mn}

\newif\ifAMStwofonts
\ifoldfss
  \ifCUPmtlplainloaded \else
    \NewTextAlphabet{textbfit} {cmbxti10} {}
    \NewTextAlphabet{textbfss} {cmssbx10} {}
    \NewMathAlphabet{mathbfit} {cmbxti10} {} % for math mode
    \NewMathAlphabet{mathbfss} {cmssbx10} {} %  "   "    "
  \fi
  \ifAMStwofonts
    \ifCUPmtlplainloaded \else
      \NewSymbolFont{upmath} {eurm10}
      \NewSymbolFont{AMSa} {msam10}
      \NewMathSymbol{\upi}     {0}{upmath}{19}
      \NewMathSymbol{\umu}     {0}{upmath}{16}
      \NewMathSymbol{\upartial}{0}{upmath}{40}
      \NewMathSymbol{\leqslant}{3}{AMSa}{36}
      \NewMathSymbol{\geqslant}{3}{AMSa}{3E}

       \let\le=\leqslant
       
    \fi
  \fi
\fi % End of OFSS

\ifnfssone
  \newmathalphabet{\mathit}
  \addtoversion{normal}{\mathit}{cmr}{m}{it}
  \addtoversion{bold}{\mathit}{cmr}{bx}{it}
  \newmathalphabet{\mathbfit} % math mode version of \textbfit{..}
  \addtoversion{normal}{\mathbfit}{cmr}{bx}{it}
  \addtoversion{bold}{\mathbfit}{cmr}{bx}{it}
  \newmathalphabet{\mathbfss} % math mode version of \textbfss{..}
  \addtoversion{normal}{\mathbfss}{cmss}{bx}{n}
  \addtoversion{bold}{\mathbfss}{cmss}{bx}{n}
  \ifAMStwofonts
    \ifCUPmtlplainloaded \else
      \UseAMStwoboldmath
      \makeatletter
      \new@mathgroup\upmath@group
      \define@mathgroup\mv@normal\upmath@group{eur}{m}{n}
      \define@mathgroup\mv@bold\upmath@group{eur}{b}{n}
      \edef\UPM{\hexnumber\upmath@group}
      \new@mathgroup\amsa@group
      \define@mathgroup\mv@normal\amsa@group{msa}{m}{n}
      \define@mathgroup\mv@bold\amsa@group{msa}{m}{n}
      \edef\AMSa{\hexnumber\amsa@group}
      \makeatother
      \mathchardef\upi="0\UPM19
      \mathchardef\umu="0\UPM16
      \mathchardef\upartial="0\UPM40
      \mathchardef\leqslant="3\AMSa36
      \mathchardef\geqslant="3\AMSa3E

       \let\le=\leqslant

    \fi
  \fi
\fi % End of NFSS release 1

\ifnfsstwo
  \DeclareMathAlphabet{\mathbfit}{OT1}{cmr}{bx}{it}
  \SetMathAlphabet\mathbfit{bold}{OT1}{cmr}{bx}{it}
  \DeclareMathAlphabet{\mathbfss}{OT1}{cmss}{bx}{n}
  \SetMathAlphabet\mathbfss{bold}{OT1}{cmss}{bx}{n}
  \ifAMStwofonts
    \ifCUPmtlplainloaded \else
      \DeclareSymbolFont{UPM}{U}{eur}{m}{n}
      \SetSymbolFont{UPM}{bold}{U}{eur}{b}{n}
      \DeclareSymbolFont{AMSa}{U}{msa}{m}{n}
      \DeclareMathSymbol{\upi}{0}{UPM}{"19}
      \DeclareMathSymbol{\umu}{0}{UPM}{"16}
      \DeclareMathSymbol{\upartial}{0}{UPM}{"40}
      \DeclareMathSymbol{\leqslant}{3}{AMSa}{"36}
      \DeclareMathSymbol{\geqslant}{3}{AMSa}{"3E}

       \let\le=\leqslant

    \fi
  \fi
\fi % End of NFSS release 2

\ifCUPmtlplainloaded \else
  \ifAMStwofonts \else % If no AMS fonts
    \def\upi{\pi}
    \def\umu{\mu}
    \def\upartial{\partial}
  \fi
\fi

\title{Theoretical insights into the RR Lyrae K-band Period-Luminosity relation}
\author[G. Bono et al.]
       {G. Bono$^1$, F. Caputo$^1$, V. Castellani$^2$, M. Marconi$^3$ and 
J. Storm$^4$ \\
        $^1$ Osservatorio Astronomico di Roma, via di Frascati 33, 00040 Monte
             Porzio Catone, Italy~~bono/caputo@coma.mporzio.astro.it\\
        $^2$ Dipartimento di Fisica, Piazza Torricelli 2, 56100 Pisa,
             Italy~~vittorio@astr18pi.difi.unipi.it\\
        $^3$ Osservatorio Astronomico di Capodimonte, Via Moiariello 16,
             80131 Napoli, Italy~~marcella@na.astro.it\\
        $^4$ Astrophysikalisches Institut Potsdam, An der Sternwarte 16,
             D-14482 Potsdam, Germany~~jstorm@aip.de}
\date{}

\begin{document}

\maketitle

\label{firstpage}

\begin{abstract}
Based on  updated nonlinear, convective pulsation models computed for  
several values of stellar mass, luminosity and metallicity, theoretical 
constraints on the $K-$band Period-Luminosity ($PL_K$) relation of 
RR Lyrae stars are presented. We show that for each given metal content  
the predicted $PL_K$ is marginally dependent on uncertainties of the 
stellar mass and/or luminosity. Then, by considering the RR Lyrae masses 
suggested by evolutionary computations for the various metallicities, 
we obtain that the predicted infrared magnitude $M_K$ over the range 
0.0001$< Z <$0.02 is given by the relation 
$M_K$=0.568$-$2.071log$P$+0.087log$Z-$0.778log$L/L_{\odot}$, with a  
rms scatter $\sigma_K$=0.032 mag. Therefore, by allowing the luminosities 
of RR Lyrae stars to vary within   the range covered by  current 
evolutionary predictions for metal-deficient (0.0001$< Z <$0.006) horizontal 
branch models, we eventually find that the infrared 
Period-Luminosity-Metallicity ($PLZ_K$) relation is 
$M_K$=0.139$-$2.071(log$P$+0.30)+0.167log$Z$, with a total intrinsic 
dispersion of $\sigma_K$=0.037 mag. 
As a consequence, the use of such a $PLZ_K$ relation should constrain 
within $\pm$0.04 mag the infrared distance modulus of field and cluster 
RR Lyrae variables, provided that accurate observations and reliable 
estimates of the metal content are available. Moreover, we show that  
the combination of $K$ and $V$ measurements can supply independent 
information on the average luminosity of RR Lyrae stars, thus yielding 
tight constraints on the input physics of stellar evolution computations.
Finally, for  globular clusters with a sizable sample of first overtone 
(RR$_c$) variables, the reddening can be estimated by using the $PLZ_K$ 
relation together with the predicted $M_V-$log$P$ relation at the blue 
edge of the instability strip (Caputo et al. 2000).
\end{abstract}


\section{Introduction}

Together with Classical Cepheids, RR Lyrae stars are popular 
standard candles for estimating stellar distances in the Milky Way 
and to Local Group galaxies. 
Cepheids are the typical Population I standard candles, while    
RR Lyraes are generally observed in globular clusters (GCs), 
thus providing a widely adopted method to calibrate the luminosity 
of the main-sequence turn-off and  to constrain the age of these 
very old stellar systems. 

In the last few years distances based on the RR Lyrae properties 
uncorked a flourishing literature. However in spite of these efforts, 
we are still facing the problem that distance determinations based 
on different empirical methods present discrepancies that are larger 
than the estimated intrinsic errors (see e.g. Carretta et al. 2000), 
thus suggesting the presence of poorly known systematic errors.
On the other hand, theoretical predictions concerning the horizontal 
branch (HB) luminosity at the RR Lyrae gap are affected by uncertainties 
in the input physics of stellar models that do not allow us to nail down 
accurate estimates of the absolute magnitude $M_V$(RR) of RR Lyrae stars 
(Cassisi et al. 1998; Caputo 1998; Castellani \& Degl'Innocenti 1999; 
Caputo et al. 2000).

The pros and cons of the various empirical methods currently adopted to 
determine RR Lyrae distances both in the field and in GCs have been 
widely discussed in the literature and they are not further addressed here.
However, it is already known that near infrared (NIR) observations, 
and in particular $K-$ band photometry, can overcome several deceptive
errors affecting the RR Lyrae distance scale.
As a matter of fact, $K$ magnitudes present some indisputable advantages 
when compared with usual $B$ or $V$ magnitudes:
i) a much smaller dependence on interstellar extinction,  
ii) a smaller dependence on stellar metallicity, and finally 
iii) smaller pulsational amplitudes which allow us to reach a superior 
precision on time averaged magnitudes with a relatively small number of 
observations.

The NIR observational scenario for RR Lyrae stars has been further enriched 
by the empirical evidence brought out by Longmore, Fernley \& Jameson
(1986) and by Longmore et al. (1990, hereinafter L90) who found that 
cluster RR Lyraes do obey a rather precise $PL_K$ relation.
On the basis of RR Lyrae $K$ photometry in eight Galactic GCs,  
L90 found that the intrinsic scatter of individual $PL_K$ relations 
was dominated by observational errors, as well as that all the 
relations have a slope $dM_K/d \log P\sim-$2.2, within the empirical 
uncertainties. The occurrence of a $PL$ relation much steeper in the 
$K$ than in the $V-$band is obviously due to the fact that $K$ magnitudes 
present a stronger dependence on the effective temperature than the optical 
ones (see also Nemec, Nemec \& Lutz 1994). As far as the  intrinsic 
accuracy of the $PL_K$ relation is concerned, it might be taken as 
evidence that the scatter of stellar masses and/or luminosities 
among RR Lyrae stars marginally affects the correlation between 
$K$ magnitudes and periods (L90).

To calibrate the empirical $PL_K$ relations, L90 provided three
zero-points according to the absolute distances of field RR Lyrae 
stars, as measured with different Baade-Wesselink methods. In all 
cases, a mild dependence on metallicity was suggested, with the 
[Fe/H] coefficient ranging from  0.04 to 0.08 mag dex$^{-1}$. 
However, L90 also found that calibrations with the IR flux method 
(Fernley et al. 1989; Skillen et al. 1989) give RR Lyrae luminosities  
$\sim$0.14 mag brighter than  predicted by the IR Baade-Wesselink method 
(Jones, Carney, \& Latham 1988, hereinafter JCL88; Liu \& Janes 1990, 
hereinafter LJ90). 
More recently, Jones et al. (1992, hereinafter J92) presented a detailed 
analysis of 18 field RR Lyrae stars with distances estimated according 
to the IR surface brightness version of the Baade-Wesselink method and 
with  metallicities ranging from solar down to [Fe/H]=$-$2.2. 
The authors concluded that the $PL_K$ relation appears largely 
insensitive to the stellar metal abundance.

Summarizing, the most recent empirical metal-free $PL_K$ relations 
available in the literature are those provided by J92  
$$M_K=-0.88(\pm0.06)-2.33(\pm0.20)\log P$$
\noindent 
and by Skillen et al. (1993)
$$M_K=-1.07(\pm0.10)-2.95(\pm0.10)\log P,$$
\noindent
while, if the metallicity 
term is taken into account, the empirical period-luminosity-metallicity 
($PLZ_K$) relations are 
$$M_K=-0.647-1.72 \log P+0.04\ [Fe/H],$$
$$M_K=-0.76-2.257 \log P+0.08\ [Fe/H],$$
\noindent
and
$$M_K=-0.72(\pm0.11)-2.03(\pm0.27)\log P+0.06(\pm 0.04) [Fe/H]$$
\noindent
by L90, LJ90 and J92, respectively. 
We add that 
the absolute calibration suggested by L90 according to 
the IR flux method distances of two field RR Lyrae stars yields   
quite brighter magnitudes, namely
$$M_{K,-0.30}=-0.24+0.06\ [Fe/H],$$
\noindent
where $M_{K,-0.30}$ is the absolute infrared magnitude at 
log$P$=$-$0.30.

At variance with such a large body of observational work, we still lack 
a detailed theoretical study of the physical scenario underlying the 
$PL_K$ relation.
In this paper, we wish to investigate this topic on the basis 
of an extensive grid of nonlinear, convective pulsation models constructed 
by adopting chemical compositions, luminosities and stellar masses typical
of field and cluster RR Lyrae variables.
In Section 2 we present the adopted sets of models and discuss the
dependence of the predicted $K-$magnitudes on period and metallicity. 
The theoretical $PLZ_K$ relation is discussed in the same section, 
together with the comparison with current
empirical data for RR Lyrae stars in M3. Final remarks and future
developments close the paper.

\begin{figure}
%\vbox to 100mm{ 
\psfig{figure=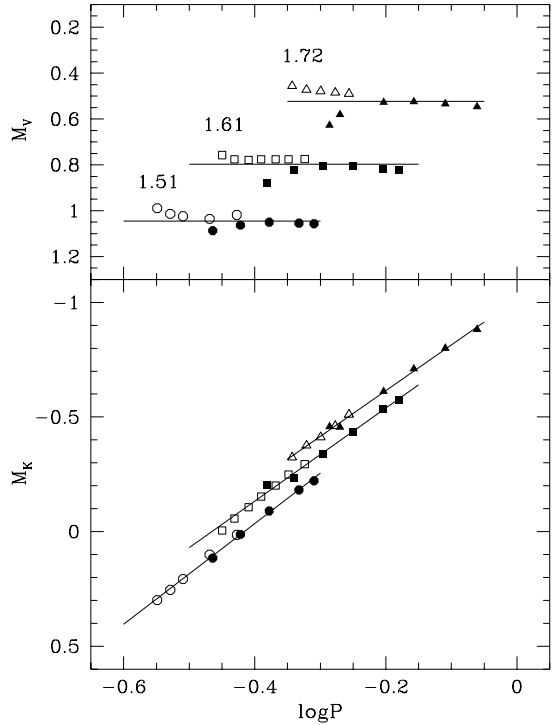,height=10cm}
\caption{Top panel: predicted $M_V$ magnitudes 
for fundamental (filled symbols)
and first overtone (open symbols) pulsators at fixed mass (0.65$M_{\odot}$)
and chemical composition ($Y$=0.24, $Z$=0.001), but with three different
luminosities (see labeled values).
The models cover the entire pulsation region, from the blue to
the red edge of the instability strip and the periods of FO pulsators are  
fundamentalised. The solid lines display the mean visual magnitude of the 
pulsators at fixed luminosity. Bottom panel: predicted $M_K$ magnitudes 
for the same pulsators. The solid lines show the $PL_K$ relations at fixed  
luminosity.}
%\vfil} 
\end{figure}

\begin{figure}
%\vbox to 100mm{ 
\psfig{figure=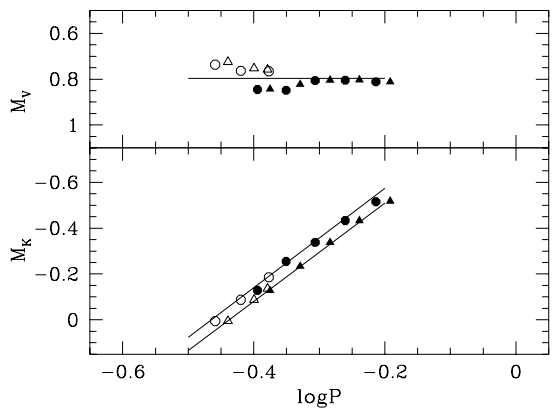,height=10cm}
\caption{Same as in Fig. 1, but with fixed luminosity
(log$L/L_{\odot}$=1.61) and two different mass values, namely 
0.625$M_{\odot}$ (triangles) and 0.675$M_{\odot}$ (circles).}
%\vfil} 
\end{figure}

\section{Predicted $PLZ_K$ relation}

As quoted before, empirical studies of the $PL_K$ relation 
suggest that both the evolutionary (i.e., luminosity) effects and 
the spread of stellar masses inside the instability strip marginally 
affect that relation. To assess the plausibility of such a scenario, 
we computed three sequences of pulsating models for fixed mass 
($0.65M_{\odot}$) and chemical composition ($Y$=0.24, $Z$=0.001),
but with three different luminosities (log$L/L_{\odot}$=1.51, 1.61, 1.72) 
and effective temperatures ranging from the blue to the red edge 
of the instability region. The theoretical framework adopted to 
construct the nonlinear convective models has been presented in 
a series of papers (Bono \& Stellingwerf 1994; Bono, Caputo, \& Marconi 1995; 
Bono et al. 1997b; Caputo et al. 2000) and it is not further discussed here.
The predicted luminosities have been transformed into the observational
plane by adopting the bolometric corrections and the color-temperature
relations provided by Castelli, Gratton, \& Kurucz (1997a,b). 

Figure 1 shows the predicted  $M_V$ (top panel) and $M_K$ (bottom panel)
magnitudes for fundamental (F) and first overtone (FO) pulsators, 
as a function of the pulsation period. Note 
that the FO models have been fundamentalized by adding 0.127
to their logarithmic period, while the predicted magnitudes are  
magnitude-weighted means over the pulsation cycle. It is worth noting that 
mean magnitudes can be derived by adopting two different methods, 
i.e. magnitude-weighted $(m_j)$ or intensity-weighted $<m_j>$. 
However, the NIR amplitudes are relatively small and therefore the 
discrepancy between synthetic $(K)$ and $<K>$ magnitudes is less than 
0.01 mag for both F and FO pulsators, in close agreement with empirical 
studies (see e.g. L90 and Fernley 1993). Concerning the visual band, 
the difference $(V)-<V>$, as well as the difference 
between static\footnote{The magnitude such a star would have were it not 
pulsating.} and mean magnitudes, 
depend on the symmetry of the light curve, generally   
increasing as the amplitude of pulsation increases.   
As shown by Bono, Caputo \& Stellingwerf (1995), the 
differences for the first-overtone mode are always 
smaller than for fundamental mode, 
and among F pulsators the discrepancies are significantly 
increasing  moving from the almost sinusoidal light curves near the red edge
to the saw-tooth ones near the blue edge of the instability region. Also 
this finding agrees  
with empirical evidence (Fernley 1993). 

A glance at the data plotted in Fig. 1 brings out 
the mechanism of the $K$ band "degeneracy". For any fixed luminosity 
level, the pulsation period increases from the blue to the red edge 
of the instability region and the $M_V$ magnitude remains, as expected, 
almost constant, except for some faster fundamental pulsators  
appearing fainter according to the light curve effect 
already discussed in Bono et al. (1995).
At any fixed luminosity the dependence of $M_V$ on period is
$dM_V/d \log P\approx-0.14$, whereas the dependence of $M_K$ is steeper,
and indeed $dM_K/d \log P\approx-2.1$. This effect is due to fact that
when moving from the blue to the red edge of the instability strip the
period increases and the pulsators become brighter, due to the strong
dependence of the K-band bolometric correction on effective temperature.
An increase in the luminosity, for any fixed effective temperature,
causes an increase in the period according to $d \log P/d \log L\sim$0.84,
while the visual magnitude closely follows the expected shift in luminosity
as $dM_V/d \log L\sim-$2.5. On the other hand, a change in the luminosity
level causes a variation in both $K$ magnitudes and periods in such a way
that the $PL_K$ relation of bright pulsators fairly match the relation
for the fainter ones (see bottom panel of Fig. 1).

Figure 2 shows $M_V$ (top panel) and $M_K$ (bottom panel) magnitudes 
for models with the quoted chemical composition, fixed luminosity 
(log$L/L_{\odot}$=1.61), but with two different mass values 
(0.625 and 0.675$M_{\odot}$) taken as a reasonable estimate of the 
expected mass dispersion among RR Lyrae stars with log$Z\sim -$3.0. 
We find that the $PL_K$ relation is only marginally affected by 
the spread in the stellar mass, and indeed a variation of 8 percent  
introduces, at fixed period, a variation in $M_K$ of $\sim$0.06 mag.

Based on the results presented in Fig. 1 and Fig. 2, we find that for 
a given chemical composition the effect on the $PL_K$ of a dispersion 
in the stellar mass of $\pm$4\% is almost negligible ($\sim\pm$0.03 mag), 
and the effect caused by a dispersion in luminosity is significantly 
smaller than in the visual relation. 
Even if we account for a dispersion as large as $\Delta \log L=\pm 0.1$,
the $PL_K$ relation would give the absolute $K$ magnitude 
(i.e. the distance modulus) with an accuracy better than 
$\pm$0.1 mag, which is substantially smaller than the analogous 
dispersion in the visual band, i.e. $\Delta M_V\sim \pm$0.25 mag. 

Theoretical predictions on the $PL_K$ relation do depend on the
adopted bolometric corrections and color-temperature relations.
Recent detailed investigations suggest that predictions based on
current atmosphere models are in good agreement with empirical
data for both dwarf and giant stars (Bessell, Castelli, \& Plez 1998).
The same outcome does not apply to RR Lyrae stars, and indeed
discrepancies of the order of 100-200 K have been found between
temperature estimates based on optical and NIR colors (Clementini
et al. 1995; Cacciari et al. 2000). Fortunately enough, the
V-K colors seem only marginally affected by systematic uncertainties,
and indeed Clementini et al. found that predicted temperatures are,
at fixed V-K, on average 50 K cooler than observed ones.
To estimate the impact of this uncertainty on $M_K$ magnitudes we
assumed an error of $\pm 50$ K on the temperature and we found
that predicted V-K colors for both metal-poor and metal-rich structures
change by $\pm 0.03$ mag. As a consequence, the uncertainty introduced
by the (V-K)-temperature calibration on $M_K$ seems negligible when
compared with evolutionary effects. 

To investigate in more detail the NIR period-luminosity 
relation, theoretical evaluations 
have been extended to the metallicity range $0.0001 \le Z \le 0.02$, 
by adopting for each given value of $Z$ a helium content fixed according 
to a Galactic helium-to-metal enrichment ratio $\Delta Y/\Delta Z\sim$2.5. 
For each  chemical composition we adopted the mass of the  
Zero Age Horizontal Branch (ZAHB) model (Bono et al. 1997a; 
Cassisi et al. 1998) located in the middle of the instability strip  
(log$T_e\sim$3.85), while several luminosity levels have been explored 
both to cover the different theoretical ZAHB luminosities predicted in 
the recent literature (Caputo et al. 2000, and references therein) 
and to account for the expected evolutionary effects. 
Table 1 summarizes the input parameters of the computed pulsating 
models, while Table 2 and Table 3 give the details of fundamental 
and first overtone pulsators, respectively\footnote{Table 2 and 
Table 3 list magnitude-weighted quantities over the 
pulsation cycle. Intensity-weighted magnitudes can be requested 
to marcella@na.astro.it.}.

\begin{table}
\centering
\caption[]{Input parameters for the pulsation models.\label{tab1}}
\begin{tabular}{llcr}
 $Z$  & $Y$ & $M/M_{\odot}$ & log $L/L_{\odot}$ \\
0.0001 & 0.24  &  0.75 & 1.72, 1.81   \\
0.0004 & 0.24  &  0.70 & 1.61, 1.72, 1.81   \\
0.001  & 0.24  &  0.65 & 1.51, 1.61, 1.72 \\
0.006  & 0.255 &  0.58 & 1.55, 1.65 \\
0.01   & 0.255 &  0.58 & 1.51, 1.57 \\
0.02   & 0.28  &  0.53 & 1.41, 1.51, 1.61 \\
\end{tabular}
\end{table}

\begin{figure}
%\vbox to 100mm{ 
\psfig{figure=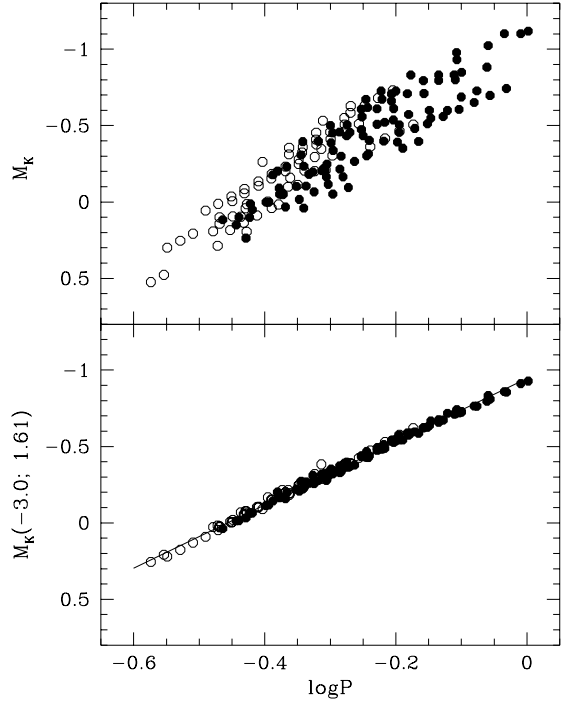,height=10cm}
\caption{Top panel: predicted $M_K$ magnitudes of F (filled circles)
and FO pulsators (open circles) as a function of period (data in Table 2
and Table 3, respectively). Bottom panel: same as the top panel, but 
with the
$M_K$ magnitudes scaled to the same metallicity ($Z$=0.001) and
luminosity (log$L/L_{\odot}$=1.61). The solid line shows the relation
predicted by eq. (1). In both panels the periods of FO pulsators are
fundamentalised.}
%\vfil} 
\end{figure}

\begin{table*}
\centering
\caption[]{Periods and absolute mean magnitudes (magnitude-weighted) 
for fundamental pulsators at
different metallicities, luminosities, and effective temperatures.\label{tab2}} 
\begin{tabular}{ccccrcccccrc}
 $Z$ & log $L/L_{\odot}$ & $T_e$ & $P$ & $M_K$ & $M_V$ &   $Z$ & log $L/L_{\odot}$ & $T_e$  & $P$ & $M_K$ & $M_V$ \\
     &                   & (K)   &  (d)& (mag) & (mag) &       &                   & (K)    & (d) & (mag) & (mag) \\
 0.0001 &  1.72 &  6900  &  0.4551    &  -0.395  &  0.606   &  0.001 &   1.72 &
 6100  &   0.7776   &  -0.800  &  0.533                \\  
 0.0001 &  1.72 &  6700	 &  0.5014    &	 -0.501  &  0.583   &  0.001 &   1.72 &  5900  &   0.8690   &  -0.882  &  0.546                \\  
 0.0001 &  1.72 &  6500	 &  0.5586    &	 -0.607  &  0.565   &  0.006 &   1.55 &  7100  &   0.3636   &   0.099  &  0.972              \\
 0.0001 &  1.72 &  6300	 &   0.6215   &	 -0.707  &  0.563   &  0.006 &   1.55 &  7000  &   0.3810   &   0.050  &  0.983              \\
 0.0001 &  1.72 &  6100	 &   0.6950   &	 -0.794  &  0.571   &  0.006 &   1.55 &  6900  &   0.3997   &   0.001  &  0.978               \\
 0.0001 &  1.72 &  6000	 &   0.7331   &	 -0.831  &  0.575   &  0.006 &   1.55 &  6800  &   0.4195   &  -0.051  &  0.968              \\
 0.0001 &  1.81 &  6800	 &   0.5679   &	 -0.672  &  0.399   &  0.006 &   1.55 &  6500  &   0.4865   &  -0.205 &  0.929               \\
 0.0001 &  1.81 &  6700	 &   0.5990   &	 -0.726  &  0.384   &  0.006 &   1.55 &  6100  &   0.6047   &  -0.398  &  0.924              \\
 0.0001 &  1.81 &  6500	 &   0.6649   &	 -0.831  &  0.356   &  0.006 &   1.55 &  5900  &   0.6741   &  -0.481 &  0.934              \\
 0.0001 &  1.81 &  6200	 &   0.7807   &	 -0.978  &  0.343   &  0.006 &   1.65 &  7000  &   0.4713   &  -0.196  &  0.690              \\
 0.0001 &  1.81 &  5900	 &   0.9237   &	 -0.110  &  0.357   &  0.006 &   1.65 &  6900  &   0.4955   &  -0.248  &  0.689             \\
 0.0004 &  1.61 &  6800	 &   0.4096   &	 -0.178  &  0.859   &  0.006 &   1.65 &  6800  &   0.5207   &  -0.299  &  0.679              \\
 0.0004 &  1.61 &  6700	 &   0.4300   &	 -0.231  &  0.844  &  0.006 &   1.65 &  6600  &   0.5747   &  -0.403  &  0.658              \\
 0.0004 &  1.61 &  6600	 &   0.4522   &	 -0.307  &  0.854   &  0.006 &   1.65 &  6400  &   0.6387   &  -0.505  &  0.648               \\
 0.0004 &  1.61 &  6400	 &   0.5024   &	 -0.387  &  0.826   &  0.006 &   1.65 &  6200  &   0.7098   &  -0.600  &  0.654              \\
 0.0004 &  1.61 &  6300	 &   0.5307   &	 -0.433  &  0.828   &  0.006 &   1.65 &  6000  &   0.7939   &  -0.687  &  0.664              \\
 0.0004 &  1.61 &  6200	 &   0.5592   &	 -0.474  &  0.831  &  0.006 &   1.65 &  5900  &   0.8387   &  -0.727  &  0.670             \\
 0.0004 &  1.61 &  6100	 &   0.5907   &	 -0.509  &  0.830   &  0.01 &    1.51 &  7000  &   0.3602   &   0.151  &  1.056              \\
 0.0004 &  1.61 &  6050	 &   0.6067   &	 -0.522  &  0.827  &  0.01 &    1.51 &  6900  &   0.3774   &   0.101  &  1.062              \\
 0.0004 &  1.72 &  6900	 &   0.4805   &	 -0.398  &  0.608   &  0.01 &    1.51 &  6500  &   0.4603   &  -0.104  &  1.023              \\
 0.0004 &  1.72 &  6800	 &   0.5055   &	 -0.451  &  0.597   &  0.01 &    1.51 &  6100  &   0.5692   &  -0.303  &  1.019              \\
 0.0004 &  1.72 &  6700	 &   0.5317   &	 -0.505  &  0.583  &  0.01 &    1.51 &  5900  &   0.6344   &  -0.392  &  1.030              \\
 0.0004 &  1.72 &  6600	 &   0.5601   &	 -0.558  &  0.569   &  0.01 &    1.57 &  7000  &   0.4036   &   0.000  &  0.882              \\
 0.0004 &  1.72 &  6500	 &   0.5904   &	 -0.610 &  0.556  &  0.01 &    1.57 &  6900  &   0.4246   &  -0.051  &  0.904              \\
 0.0004 &  1.72 &  6400	 &   0.6212   &	 -0.661  &  0.548  &  0.01 &    1.57 &  6800  &   0.4447   &  -0.102  &  0.901              \\
 0.0004 &  1.72 &  6300	 &   0.6570   &	 -0.709  &  0.546   &  0.01 &    1.57 &  6600  &   0.4919   &  -0.205  &  0.881              \\
 0.0004 &  1.72 &  6100	 &   0.7332   &	 -0.795  &  0.553   &  0.01 &    1.57 &  6100  &   0.6384   &  -0.456  &  0.865              \\
 0.0004 &  1.72 &  6000	 &   0.7746   &	 -0.833  &  0.556   &  0.01 &    1.57 &  5900  &   0.7134   &  -0.548  &  0.877              \\
 0.0004 &  1.72 &  5950	 &   0.7946   &	 -0.848  &  0.557   &  0.01 &    1.57 &  5800  &   0.7565   &  -0.599  &  0.887              \\
 0.0004 &  1.81 &  6900	 &   0.5712   &	 -0.619  &  0.383   &  0.02 &    1.41 &  6700  &  0.3726    &  0.237   &  1.235              \\
 0.0004 &  1.81 &  6800	 &   0.6009   &	 -0.672  &  0.371   &  0.02 &    1.41 &  6300  &  0.4565    &  0.040   &  1.227              \\
 0.0004 &  1.81 &  6700	 &   0.6308   &	 -0.726  &  0.355   &  0.02 &    1.41 &  6100  &  0.5056    & -0.052  &  1.236             \\
 0.0004 &  1.81 &  6300	 &   0.7819   &	 -0.932  &  0.317  &  0.02 &    1.41 &  6000  &  0.5342    & -0.095   &  1.241              \\
 0.0004 &  1.81 &  6100	 &   0.8728   &	 -0.102  &  0.324   &  0.02 &    1.51 &  6800  &  0.4282    &  0.033   &  0.994              \\
 0.0004 &  1.81 &  5900	 &   0.9781   &	 -0.110  &  0.335   &  0.02 &    1.51 &  6700  &  0.4494    & -0.017   &  0.987              \\
 0.0004 &  1.81 &  5850	 &   0.1005   &	 -0.112  &  0.338   &  0.02 &    1.51 &  6600  &  0.4733    & -0.066   &  0.981              \\
 0.001 &   1.51 &  6900	 &   0.3434   &  0.116	  &  1.087  &   0.02 &    1.51 &  6500  &  0.4975    & -0.115   &  0.977             \\
 0.001 &   1.51 &  6700	 &   0.3785   &	 0.012	  &  1.063  &   0.02 &    1.51 &  6400  &  0.5239    & -0.165   &  0.977             \\
 0.001 &   1.51 &  6500	 &   0.4187   &	 -0.090  &  1.050  &   0.02 &    1.51 &  6000  &  0.6463    & -0.351   &  0.994             \\
 0.001 &   1.51 &  6300	 &   0.4650   &	 -0.182  &  1.054  &   0.02 &    1.51 &  5900  &  0.6835    & -0.396   &  1.002             \\
 0.001 &   1.51 &  6200	 &   0.4902   &	 -0.222  &  1.057  &   0.02 &    1.61 &  6900  &  0.4944    & -0.165   &  0.745             \\  
 0.001 &   1.61 &  6900	 &   0.4163   &	 -0.202  &  0.877   &  0.02 &    1.61 &  6800  &  0.5182    & -0.215   &  0.742              \\ 
 0.001 &   1.61 &  6700  &   0.4567   &  -0.234  &  0.824   &  0.02 &    1.61 &  6700  &  0.5456    & -0.265   &  0.737              \\ 
 0.001 &   1.61 &  6500  &   0.5061   &  -0.338  &  0.805   &  0.02 &    1.61 &
 6600  &  0.5731    & -0.315   &  0.733              \\ 
 0.001 &   1.61 &  6300  &   0.5626   &  -0.433  &  0.804   &  0.02 &    1.61 &
 6200  &  0.7052    & -0.512   &  0.733             \\
 0.001 &   1.61 &  6100  &   0.6254   &  -0.537  &  0.820   &  0.02 &    1.61 &  6100  &  0.7452    & -0.559   &  0.739              \\ 
 0.001 &   1.61 &  6000  &   0.6598   &  -0.573  &  0.823   &  0.02 &    1.61 &  6000  &  0.7877    & -0.606   &  0.745              \\
 0.001 &   1.72 &  7000	 &   0.5178   &	 -0.458  &  0.627   &   0.02 &    1.61 &  5900 &  0.8306    & -0.651   &  0.753            \\ 
 0.001 &   1.72 &  6800	 &   0.5369   &	 -0.456  &  0.581   &   0.02 &    1.61 &  5800 &  0.8783    & -0.697   &  0.763            \\ 
 0.001 &   1.72 &  6500  &   0.6260   &  -0.611  &  0.527   &   0.02 &    1.61 &  5700 &  0.9305    & -0.743  &  0.775             \\
 0.001 &   1.72 &  6300  &   0.6964   &  -0.710  &  0.524  &  \ldots&  \ldots &  \ldots&      \ldots&     \ldots&  \ldots          \\ 
\end{tabular}  						   
\end{table*}

\begin{table*}
\centering
\caption[]{Same as Table 2, but for first overtone pulsators.\label{tab3}} 
\begin{tabular}{ccccrcccccrc}
 $Z$ & log $L/L_{\odot}$ & $T_e$ & $P$ & $M_K$ & $M_V$ &   $Z$ & log $L/L_{\odot}$ & $T_e$  & $P$ & $M_K$ & $M_V$ \\
     &                   & (K)   &  (d)& (mag) & (mag) &       &                   & (K)    & (d) & (mag) & (mag) \\
0.0001 &  1.72  &    7200    &    0.2961    &   -0.263    &   0.469   &  0.001   & 1.61   &   6600     &   0.3558     &  -0.295    &  0.775  \\
0.0001 &  1.72  &    7000    &    0.3253    &   -0.355	   &   0.517   &  0.001  &  1.72   &   7100     &   0.3404     &  -0.324     &  0.456 \\
0.0001 &  1.72  &    6800    &    0.3577    &   -0.454	   &   0.523   &  0.001  &  1.72   &   7000     &   0.3580     &  -0.375     &  0.472 \\
0.0001 &  1.72  &    6600    &    0.3948    &   -0.551	   &   0.523   &  0.001  &  1.72   &   6900     &   0.3762     &  -0.412     &  0.478\\
0.0001 &  1.81  &    7100    &    0.3661    &   -0.531	   &   0.257   &  0.001  &  1.72   &   6800     &   0.3959     &  -0.461     &  0.485\\
0.0001 &  1.81  &    6900    &    0.4038    &   -0.628	   &   0.286   &  0.001 &   1.72   &   6700     &   0.4156     &  -0.511     &  0.490\\
0.0004 &  1.61  &    7300    &    0.2425    &   0.056 	   &   0.760   &  0.006 &   1.55  &    7200    &    0.2548    &   0.143     &   0.842\\
0.0004 &  1.61  &    7200    &    0.2534    &   0.012	   &   0.805   &  0.006  &  1.55  &    7100    &    0.2668    &   0.095     &   0.859\\
0.0004 &  1.61  &    7100    &    0.2655    &   -0.036	   &   0.812   &  0.006  &  1.55  &    6700    &    0.3242    &   -0.100    &   0.877 \\
0.0004 &  1.61  &    7000    &    0.2780    &   -0.085	   &   0.810   &  0.006  &  1.65  &    7100    &    0.3264    &   -0.158    &   0.585\\
0.0004 &  1.61  &    6900    &    0.2919    &   -0.135	   &   0.802   &  0.006  &  1.65  &    7000    &    0.3431    &   -0.206    &   0.600 \\
0.0004 &  1.61  &    6800    &    0.3059    &   -0.184	   &   0.794  &  0.006   & 1.65  &    6900    &    0.3642    &   -0.347    &   0.626 \\
0.0004 &  1.61  &    6700    &    0.3210    &   -0.232	   &   0.790  &  0.006   & 1.65  &    6800    &    0.3786    &   -0.304    &   0.611 \\
0.0004 &  1.61  &    6600    &    0.3363    &   -0.278	   &   0.789  &  0.01 &    1.51  &    7100    &    0.2490    &    0.192    &   0.945 \\
0.0004 &  1.72  &    7100    &    0.3249    &   -0.312	   &   0.486  &  0.01 &    1.51  &    6900    &    0.2744    &    0.096    &   0.960 \\
0.0004 &  1.72  &    7000    &    0.3412    &   -0.360	   &   0.497  &  0.01 &    1.51  &    6700    &    0.3019    &   -0.002    &   0.968 \\
0.0004 &  1.72  &    6900    &    0.3573    &   -0.408	   &   0.502  &  0.01 &    1.51  &    6600    &    0.3174    &   -0.050    &   0.971 \\
0.0004 &  1.72  &    6800    &    0.3759    &   -0.458	   &   0.505  &  0.01 &    1.57  &    7100    &    0.2791    &    0.041    &   0.773 \\
0.0004 &  1.72   &   6700     &   0.3951     &  -0.507	    &  0.506  &  0.01 &    1.57  &    6600    &    0.3558    &   -0.202    &   0.815 \\
0.0004 &  1.81 	 &   7000     &   0.4033     &  -0.583	    &  0.250  &  0.02 &    1.41  &    7300    &    0.2000    &    0.524    &  1.155 \\
0.0004 &  1.81 	 &   6900     &   0.4242     &  -0.631	    &  0.263  &  0.02 &    1.41  &    7200    &    0.2095    &    0.477    &  1.171  \\
0.0004 &  1.81 	 &   6800     &   0.4442     &  -0.680	    &  0.270  &  0.02 &    1.41  &    6800    &    0.2530    &    0.287    &   1.183 \\
0.0004 &  1.81 	 &   6700     &   0.4680     &  -0.730	    &  0.277  &  0.02 &    1.41  &    6600    &    0.2800    &    0.194    &  1.186 \\
0.001 &	  1.51 	 &   7300     &   0.2119     &  0.298	    &  0.989  &  0.02 &    1.51  &    7100    &    0.2644    &   0.183     &   0.911 \\
0.001 &	  1.51  &    7200    &    0.2218    &   0.254	   &   1.014  &  0.02 &    1.51  &    7000    &    0.2773    &   0.135     &   0.925 \\
0.001 &	  1.51  &    7100    &    0.2320    &   0.207	   &   1.024  &  0.02 &    1.51  &    6900    &    0.2905    &   0.088     &   0.923 \\
0.001  &  1.51  &    6900    &    0.2547    &   0.100	   &   1.035  &  0.02 &    1.51  &    6800    &    0.3058    &   0.040     &   0.928 \\
0.001 &	  1.51  &    6700    &    0.2803    &   0.014 	   &   1.018  &  0.02 &    1.51  &    6750    &    0.3134    &   0.016     &   0.930 \\
0.001  &  1.61 	 &   7200     &   0.2663     &  -0.005	    &  0.757  &  0.02 &    1.61  &    7000    &    0.3351    &   -0.112    &   0.657 \\
0.001  &  1.61 	 &   7100     &   0.2779     &  -0.057	    &  0.776  &  0.02 &    1.61  &    6800    &    0.3696    &   -0.209    &   0.673 \\
0.001  &  1.61 	 &   7000     &   0.2921     &  -0.106	    &  0.778  &  0.02 &    1.61  &    6500    &    0.4324    &   -0.362    &   0.695 \\
0.001  &  1.61 	 &   6900     &   0.3055     &  -0.153	    &  0.776  &  0.02 &    1.61  &    6400    &    0.4555    &   -0.416    &   0.705 \\
0.001  &  1.61 	 &   6800     &   0.3211     &  -0.201	    &  0.776  &  0.02 &    1.61  &    6300    &    0.4792    &   -0.464    &   0.713 \\
0.001  &  1.61 	 &   6700     &   0.3362     &  -0.249	    &  0.775  &  0.02 &    1.61  &    6200    &    0.5030    &   -0.506    &   0.717 \\
\end{tabular}  							     
\end{table*}

\begin{figure}
%\vbox to 100mm{ 
\psfig{figure=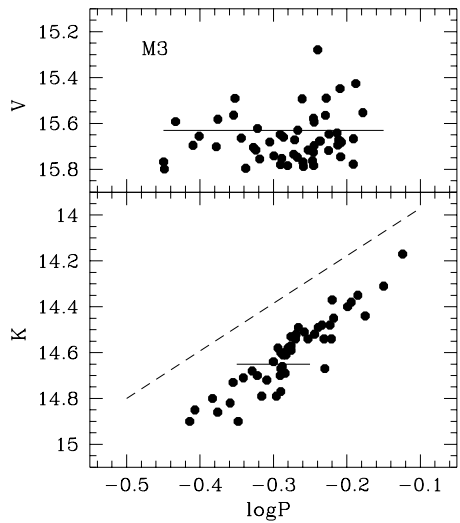,height=10cm}
\caption{Top panel: $V-$log$P$ diagram of RR Lyrae stars in M3 from 
data collected by Carretta et al. (1998). The solid line shows the 
average visual magnitude (15.65 mag). Bottom panel: $K-$log$P$ 
diagram of RR Lyrae stars in M3 from data collected by L90.  
The dashed line has the slope predicted by eq. (1), while the solid line 
is the average infrared magnitude (14.65 mag) at log $P$=$-$0.30. 
In both panels the periods of $c-$type variables are fundamentalised.} 
%\vfil} 
\end{figure}

\begin{figure}
%\vbox to 100mm{ 
\psfig{figure=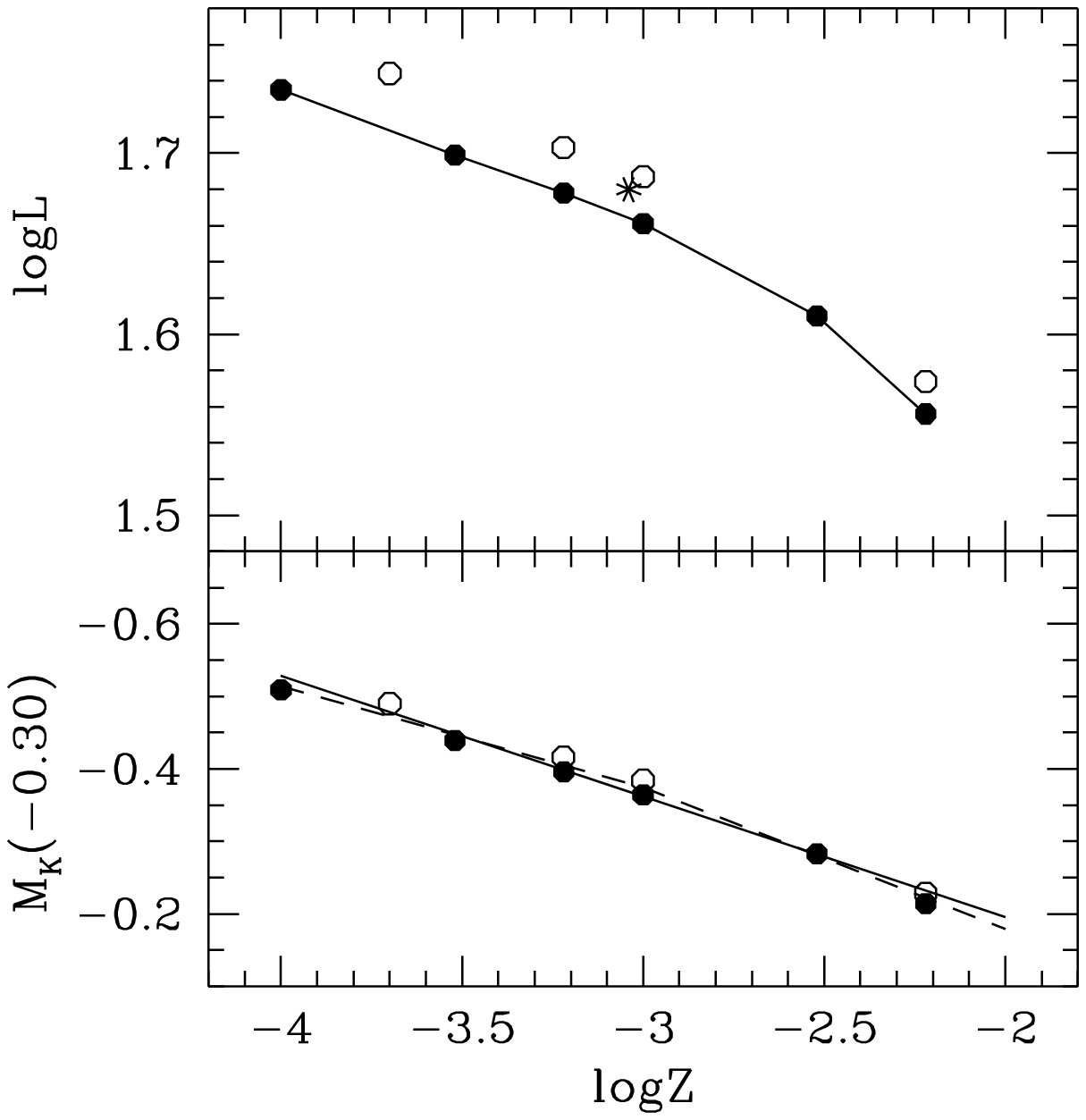,height=10cm}
\caption{Top panel: luminosity (solar units) versus metallicity  
of ZAHB models located in the middle of the RR Lyrae gap, as given by 
Bono et al. (1997a, filled circles) and Cassisi et al. 
(1998,1999, open circles). The asterisk marks the average luminosity 
of RR Lyrae stars in M3, as derived in the present paper. 
Bottom panel: predicted $K$ magnitudes at log$P$=$-0.30$ for the two 
sets of models, according to eq. (1). The solid line is  eq. (2), 
while the dashed lines refer to the $PLZ_K$ relations with log$Z\le -$3.0 
and log$Z> -$3.0.}
%\vfil} 
\end{figure}

The top panel of Fig. 3 shows the predicted $M_K$ magnitudes for the entire 
set of models plotted as a function of period. We find that $M_K$ is 
tightly correlated  with period, metallicity and luminosity according to
the relation
$$M_K = 0.568-2.071\log P+0.087\log Z-0.778\log L/L_{\odot}\eqno(1)$$
\noindent
with a rms scatter  $\sigma_K$=0.016 mag. 

The accuracy 
of eq. (1) in reproducing the predicted $M_K$ magnitudes is disclosed 
in the bottom panel of Fig. 3, 
where the magnitudes are scaled
to $Z$=0.001 and log$L/L_{\odot}$=1.61, according to eq. (1). 
Taking into consideration the uncertainty of $\sim$0.03 mag 
due to a mass dispersion by $\pm$4\% with respect to the values in Table 1, 
we eventually find that 
the total intrinsic dispersion of eq. (1) is $\pm$ 0.032 mag. 
As a result, we conclude that over the entire explored 
metallicity range a variation 
of $\Delta\log L$=$\pm$0.1 
moves the predicted $K$ magnitudes by $\pm$0.08 mag, while a 
a change in metallicity of 0.3 dex causes a change of 
$\Delta K\sim$0.03 mag. This evidence implies that RR Lyrae infrared 
magnitudes are excellent standard candles to derive distances of field 
and cluster RR Lyrae with accurate estimates of metal abundances. 
Furthermore, as we will discuss in the following, the milder  
sensitivity of the $PL_K$ relation to the stellar luminosity  
supplies the opportunity for testing 
theoretical predictions on the ZAHB luminosity.

As an example, Fig. 4 shows the observed $K-$log$P$ and $V-$log$P$ diagrams 
of RR Lyrae stars in the Galactic globular cluster M3 (data collected by 
L90 and by Carretta et al. 1998, respectively). 
The slope of the predicted $PL_K$ relation (dashed line) appears 
in reasonable agreement with observations, even though there is
a significant dispersion in the empirical data which seems to 
be dominated by observational errors, as suggested by L90.   
However, new and more accurate data are necessary to assess whether 
this dispersion is intrinsic or caused by observational errors. 
For the sake of the discussion, let us take into account  the mean 
of all the visual magnitudes, i.e. $V=$15.65$\pm$0.10 mag, and the 
averaged infrared magnitude at log$P$=$-$0.30, i.e. $K=$14.65$\pm$0.05 mag. 
By assuming for M3 an iron-to-hydrogen ratio [Fe/H]=$-$1.34 
(Carretta \& Gratton 1997), i.e. log$Z$=$-$3.04, the data in Table 2 
and Table 3 (see also the top panel of Fig. 1) 
provide the predicted averaged $M_V$ magnitude of the pulsators,  
as a function of the adopted luminosity. 
At the same time, under the same 
assumptions on metallicity and luminosity, eq. (1) yields the predicted 
$M_K$ magnitude at log$P$=$-$0.30. The absolute visual and 
infrared magnitudes are listed 
in Table 4 [column (2) and column (3), respectively], together with 
the corresponding visual ($\mu_V$) and infrared
($\mu_K$) distance moduli of the RR Lyrae stars in M3 
 [column (4) and column (5), respectively].  

Inspection of the data in Table 4 shows that an increase in the luminosity 
of $\Delta \log L$=0.21 causes an increase in the distance modulus of 
0.52 mag in the visual but only 0.16 mag in the infrared. 
Since the reddening towards M3 is negligible (Dutra \& Bica 2000) and 
can be set to zero, the average luminosity $<\log L/L_{\odot}>_{RR}$ 
of RR Lyrae stars can be found as the luminosity for which the distance 
moduli in $V$ and $K$ assume the same value 
($\mu_0=\mu_K=\mu_V$). Thus, interpolation in Table 4 leads to  
$<\log L/L_{\odot}>_{RR}$=1.68$\pm$0.05 for a mean distance modulus 
$\mu_K$=15.03$\pm$0.07 mag,  where 
the error on the distance modulus also includes   
the observed dispersion of infrared data. We wish to notice that 
this value of the M3 distance modulus is in close agreement with
that ($\mu_V=15.00\pm$0.07 mag) derived by Caputo et al. (2000) 
on the basis
of the comparison between the observed distribution in the $V-$log$P$
plane of M3 $c-$type variables and the predicted blue edge for first
overtone pulsators, as given by the same pulsating models adopted in the
present paper. This agreement confirms the internal consistency of 
our pulsational scenario, as well as that the reddening towards M3 is 
almost zero. 

In this context it is 
worth emphasizing that  
the sensitivity of $<\log L/L_{\odot}>_{RR}$ to errors in 
the reddening determination has a rather negligible 
effect for the true distance modulus derived by infrared observations. 
As a matter of fact, if we assume for M3 a reddening $E(B-V)\sim$0.05 mag. 
With $A_V/E(B-V)$=3.1 and $A_K/E(V-K)$=0.13 (see Cardelli, 
Clayton \& Mathis 1989),  
the apparent visual 
and infrared moduli in Table 4 should be decreased by 0.16 mag 
and 0.02 mag, respectively, and the condition $\mu_{0,V}=\mu_{0,K}$ 
leads now to a higher luminosity ($<\log L/L_{\odot}>_{RR}$=1.76$\pm$0.05), 
but for a quite similar true distance modulus, i.e.  
$\mu_{0,K}=15.07 \pm$0.07 mag. On a general point of view the dependence 
of $<\log L/L_{\odot}>_{RR}$ on the adopted reddening is 
$\delta \log L/\delta E(B-V)\sim$1.6 dex mag$^{-1}$, whereas the 
sensitivity of the true distance modulus based on infrared observations 
is only  $\delta \mu_{0,K}/\delta E(B-V)\sim$0.9.  
 
Finally, we wish to notice 
that our distance evaluation is in close 
agreement with the M3 distance estimated by L90, i.e. 
$\mu_K =15.00\pm0.04\pm0.15$,  where the uncertainties refer to errors in 
the zero-point and in the slope, respectively. As a consequence, our 
theoretical approach to the $PLZ_K$ relation seems to 
support the absolute calibration based on the 
IR flux method. A detailed comparison between individual cluster 
distances based on empirical methods and theoretical predictions, as well 
as a thorough analysis of the error budget, will be addressed in a 
forthcoming paper.  
 
\begin{table}
\centering
\caption[]{Predicted $M_V$ and $M_K$  (at log$P$=$-$0.30) magnitudes for
pulsators with log$Z$=$-3.04$, as a function of the adopted luminosity 
(see text). 
The 
visual ($\mu_V$) and infrared  ($\mu_K$) distance moduli of RR Lyrae
stars in M3 are
derived according to the observed averaged magnitudes $V$=15.65 mag
and $K$($-$0.30)=14.65 mag. The last column gives the difference between 
visual and infrared distance moduli.\label{tab4}}
\begin{tabular}{cccccc}
log $L/L_{\odot}$ & $M_V$ & $M_{K,-0.30}$ & $\mu_V$ & $\mu_K$ & 
$\Delta \mu$      \\
 1.51  & 1.045 & -0.250   & 14.605  & 14.900  & $-$0.295\\
 1.61  & 0.797 & -0.328   & 14.853  & 14.978  & $-$0.125\\
 1.72  & 0.523 & -0.413   & 15.127  & 15.063  & $+$0.064\\
\end{tabular}
\end{table}

\section{Discussion}

In the previous section it has been shown that, if one relies on theoretical 
pulsation results, the predicted $PLZ_K$ relation of RR Lyrae stars can 
supply accurate distance estimates to globular clusters with reliable 
infrared observations. In fact, this relation appears only marginally
dependent on the exact value of the luminosity as 
well as on current uncertainties on the iron-to-hydrogen content. Moreover, 
by combining $K$ and $V$ magnitudes one can derive independent constraints 
on the average bolometric magnitude of RR Lyrae variables. 

In this context, we show in Fig. 5 (top panel) the log$L$(ZAHB)$-$log$Z$ 
relations  based on model computations 
by Bono et al. (1997a, filled circles) and by 
Cassisi et al. (1998, 1999, open circles), which cover current uncertainties 
on the ZAHB luminosity, together with the average luminosity of M3 RR Lyrae 
stars (asterisk), as derived in the previous section. We find that the ZAHB 
luminosities predicted by Bono et al. (1997a) agree quite well with the 
expected average overluminosity ($\delta$log$L\sim$0.04) 
of actual RR Lyrae stars when compared with the ZAHB 
luminosity. On the other hand, the HB models computed by 
Cassisi et al. (1998, 1999) using the most updated physics do predict 
ZAHB luminosities that are systematically too bright. This seems to 
strengthen the evidence based on independent estimates (Bono et al. 2000;
Caputo et al. 2000) that the most updated physics could be 
still affected by systematic uncertainties. 
 
However, the bottom panel of Fig. 5 shows that the current uncertainty on  
ZAHB luminosity has a rather weak effect on $K$ magnitudes at fixed period. 
Then, by adopting 
the two log$L$(ZAHB)$-$log$Z$ calibrations plotted in Fig. 5 as a 
measure of the luminosity dispersion among actual variables with 
a given metallicity, 
we find that the predicted $K$ magnitude of RR Lyrae stars 
with metal content in the range 
0.0001$< Z <$0.006 is correlated with period and metallicity 
as 

$$M_K  = 0.139-2.071(\log P+0.30)+0.167\log Z,\eqno(2)$$
\noindent
with a total intrinsic dispersion of $\sigma_K$=0.037 mag, including 
the above uncertainty on luminosities and mass variations by $\pm$4\% 
on the values listed in Table 1. If we also wish to account  
for the nonlinearity of the log$L$(ZAHB)$-$log$Z$ 
relationship and for the change of the slope at log$Z\sim-$3.0, then  
the least-squares solutions for models with log$Z\le -$3.0 and log$Z > -$3.0 
supply  
$$M_K = 0.058-2.071(\log P+0.30)+0.141\log Z\eqno(3)$$
\noindent 
and 
$$M_K  = 0.213-2.071(\log P+0.30)+0.196\log Z\eqno(4)$$
\noindent
respectively, with $\sigma_K$=0.035 mag. 

It is worth  mentioning that for globular clusters with a sizable sample 
of $c-$type variables, the visual distance modulus may be derived by 
matching the observed RR$_c$ distribution in the $V$-log$P$ plane with  
the predicted first overtone blue edge (FOBE). 
According to 
Caputo et al. (2000), the Period-Luminosity-Metallicity relation for 
FOBE pulsators is   
$$M_V = -0.178-2.255 \log P+0.151\log Z,\eqno(5)$$
\noindent
with a total intrinsic dispersion of $\sigma_V$=0.065 mag. 
It follows that the comparison of $K$ and $V$ magnitudes with 
theoretical predictions gives the opportunity to derive reliable 
estimates of the cluster reddening. Fortunately enough, the 
coefficients of the metallicity term in eq. (2) and eq. (5) 
are approximately the same, and therefore they cancel out any 
metallicity effect on RR Lyrae colors.  

Of course, the results presented in this paper depend on the reliability 
of the theoretical scenario concerning both pulsational models and
bolometric corrections. To quantitatively assess their 
intrinsic accuracy is obviously a difficult task, owing to the rather 
sophisticated physics involved in the modeling of stellar pulsation 
and stellar atmosphere. However, at least for the pulsation framework, 
we have shown that two quite different observables based on the 
same pulsating models, namely the $PLZ_K$
relation given in the present paper  
and the $M_V$-log$P$ relation for FOBE pulsators provided 
by Caputo et al. (2000), 
consistently provide similar values for the M3 distance modulus. 
Moreover,  in a recent paper it has been shown  that the same pulsation 
models can closely reproduce the details of the observed light curve of 
RR Lyrae stars (Bono, Castellani \& Marconi 2000). As a whole, 
these results can be taken as further evidence that the theoretical 
framework we are constructing in the last years is approaching the
behavior of actual pulsating stars.

Finally, let us note that even though $K$ observations present 
several significant advantages when compared with optical bands, 
it is limited how deep one can reach from the ground due to the 
high level of thermal background emission. 
{\it De facto} NIR data are only available for Galactic RR Lyrae stars,  
and even with the largest ground based telescopes, the observations 
can only be extended out to a few Local Group galaxies. 
However, NGST could change this picture substantially with its superior
NIR capability, and bring most Local Group galaxies within reach.
The $PLZ_K$ relation does provide a method to determine very precise
distances to these galaxies based on well understood physics.
Therefore, it will supply a very important independent check on the
Cepheid distance scale, as well as reliable calibrations of secondary
distance indicators.

{\bf Acknowledgments:} 
It is a pleasure to thank C. Cacciari and G. Clementini for useful  
discussions on current semi-empirical, color-temperature relations
for RR Lyrae stars. We also acknowledge an anonymous referee for 
his/her pertinent suggestions that improved the content of the paper. 
Financial support for this work was provided by MURST-Cofin 2000,  
under the scientific project "Stellar observables of cosmological 
relevance".

%%%%%%%%%%%%%%%%%%%%%%%%%%%%%%%%%%%%%%%%%%%%%%%%%%%%%%%%%%%%%%%%%%%%%%%%%%%%%

\end{document}